\documentclass[12pt]{article}

\usepackage{amssymb}
\usepackage{amsmath}
\usepackage{amsfonts}
\usepackage{graphicx}
\usepackage{subfigure}
\usepackage{color}
\usepackage{dcolumn}
\usepackage{hyperref}
\numberwithin{equation}{section}
\newcommand{\be}{\begin{equation}}
\newcommand{\ee}{\end{equation}}
\newcommand{\bea}{\begin{eqnarray}}
\newcommand{\eea}{\end{eqnarray}}

\begin{document}

\begin{titlepage}
\thispagestyle{empty}


\vspace{2cm}

\begin{center}
\font\titlerm=cmr10 scaled\magstep4 \font\titlei=cmmi10
scaled\magstep4 \font\titleis=cmmi7 scaled\magstep4 {
\Large{\textbf{ Diffusivities bounds in the presence of Weyl corrections}
\\}}
\vspace{1.5cm} \noindent{{
Ali Mokhtari$^{a}$\footnote{e-mail:ali.mokhtari.jazi@gmail.com }, Seyed Ali Hosseini Mansoori$^{b,c}$\footnote{e-mail:shosseini@shahroodut.ac.ir;shossein@ipm.ir}, Kazem Bitaghsir Fadafan$^{b}$\footnote{e-mail:bitaghsir@shahroodut.ac.ir}
}}\\
\vspace{0.8cm}

{\it ${}^a$ Department of Physics, Tarbiat Modares University, Tehran 14155-4838, Iran\\ }
{\it ${}^b$ Faculty of Physics, Shahrood University of Technology, P.O.Box 3619995161 Shahrood, Iran \\ }
{\it ${}^c$ School of Astronomy, Institute for Research in Fundamental Sciences (IPM) P.~O.~Box 19395-5531, Tehran, Iran\\}

\vspace*{.4cm}

\end{center}

\vskip 2em

\begin{abstract}
In this paper, we investigate the behavior of the thermoelectric DC conductivities in the presence of Weyl corrections with momentum dissipation in the incoherent limit. Moreover, we compute the butterfly velocity and study the charge and energy diffusion with broken translational symmetry. Our results show that the Weyl coupling $\gamma$, violates the bounds on the charge and energy diffusivity. It is also shown that the Weyl corrections violate the bound on the DC electrical conductivity in the incoherent limit.

\end{abstract}

\end{titlepage}

\tableofcontents

\section{Introduction}
It seems that the minimum relaxation timescale of strongly correlated materials, \emph{e.g.} strange metals, is responsible for the universal transport behavior of such systems. For example, the Ohmic resistivity of strange metals depends linearly on temperature which is in contrast with Fermi liquid metals where the Ohmic resistivity depends quadratically on temperature \cite{Sachdev:2011cs}. Based on experimental results, the strange metals in condensed matter physics are generically strongly interacting many-body quantum physics. Thus the usual perturbative approaches are not efficient to study them requiring non-perturbative new methods \cite{Bruin804,Zhang:2016ofh}.

One important new tool for studying these systems in the condensed matter is the gauge-string duality \cite{Hartnoll:2016apf}. It maps strongly correlated systems to classical gravity in higher dimensional space. With using this approach, many different aspects of such systems in the condensed matter, \emph{e.g.} transport properties of strange metals, have been studied \cite{z}. One important result is that one may study bounds on the electrical conductivity $\sigma$ and the thermal conductivity $\kappa$ in strongly correlated systems  \cite{Grozdanov:2015qia,Grozdanov:2015djs}. For example, based on the near horizon geometry of black holes, $\sigma$ has been studied in \cite{1406.4742,Donos:2015gia,Donos:2012js, Donos:2014gya, Banks:2015wha}. However, it has been shown that considering interactions of the Maxwell term with additional fields or studying the non-linear deformation of the theory leads to violation of the proposed bound \cite{Baggioli:2016oju,Fadafan:2016gmx, Baggioli:2016oqk,  Gouteraux:2016wxj, Garcia-Garcia:2016hsd, Baggioli:2016pia}. One should notice that such bounds on the transport properties should be verified experimentally and even investigate theoretically.

It is interesting that in the incoherent limit the transport is given by the diffusion of charge and energy and the microscopic mechanism of the momentum dissipation does not play any role. In this limit, the velocity of the quantum excitations $v$ and the minimum relaxation timescale $\tau$ enter a universal bound for the diffusivities as \cite{Hartnoll:2014lpa}
\begin{equation} \label{boundD}
D > v^2 \tau
\end{equation}
here, $D$ is the diffusion constant of the system. It is conjectured that this bound is related to the universal behavior of the strange metals. Recently it was proposed that the characteristic velocity of $v$ should be the butterfly velocity $v_B$ which is related to the speed of propagation of information through a quantum system \cite{Blake:2016wvh, 1604.01754}. The idea is based on the statement that the classical chaos is related to the behavior of the initial states. Recently possible constraints by quantum chaos on the hydrodynamics have been studied in \cite{Lucas:2017ibu}. With using the holography, the butterfly velocity has been computed in \cite{Roberts:2014isa} and \cite{Roberts:2016wdl}.

The proposed bound $D > v_B^2 \tau$, has been studied in simple models with momentum relaxation from holography \cite{Blake:2016wvh, 1604.01754, 1612.00849}. In the incoherent limit, the diffusion constant contains the energy diffusion constant $D_c=\frac{\sigma}{\chi}$ sector and the charge diffusion constant $D_e=\frac{\kappa}{C_{q}}$ sector where $\chi$ and $C_{q}$ are charge susceptibility and heat capacity, respectively. One should notice that $\chi$ not only depends on the horizon data but also the full bulk theory, then a violation of $D_c$ is expected. Therefore, one does not expect to find a universal behavior in the presence of new interactions. However, from the definition of the $D_e$, one concludes that it is given from the horizon behavior of the background and it is likely to observe universal behavior \cite{Blake:2017qgd}. Indeed, the violation bound of $D_c$ has been studied in \cite{Lucas:2016yfl, Baggioli:2016pia,1612.00849}. One finds the study of diffusion and butterfly velocity at finite density in \cite{Kim:2017dgz}. Butterfly velocity and thermal diffusivity in anisotropic systems have been studied in \cite{Ahn:2017kvc,Giataganas:2017koz,M3}. In the quantum region, charge diffusion constant was studied in \cite{Ling:2017jik} from holography.

It could be very interesting to understand to which extent the bound is generic. For example, in Horndeski models where one considers the scalar-tensor theories and new scalar derivatively coupled to the gravity \cite{Baggioli:2017ojd}. It is shown that in this model a subleading contribution to the thermoelectric conductivities appears which does not violate the energy diffusion and the thermal conductivity bounds. The other example is considering the Gauss-Bonnet (GB) background as higher derivative gravity. It is shown in \cite{Donos:2017oym} how the conductivity can be obtained by studying the auxiliary fluid on a curved black hole horizon. Also it was shown that in GB gravity, the thermoelectric conductivity is independent of the GB coupling \cite{Cheng:2014tya}. The universal properties of the strange metals and the electrical DC conductivity of massive N = 2 hypermultiplet fields in the GB background have been studied in \cite{Fadafan:2012hr} and \cite{AliAkbari:2010av}, respectively.

Another way to check the bound is to consider coupling of the Maxwell term to the bulk Weyl tensor. It provides a very clean test of the universality bounds on the energy diffusion at a higher order. In this case, one should analyze the bound by considering higher dimension operators in the bulk background. Such terms give interesting effects on the shear viscosity to entropy density ratio $(\frac{\eta}{s})$ \cite{0903.2834}. These terms are expected to arise in a string theory as corrections to the effective low energy field theory in a top-down approach. There are different combinations of such terms in the five-dimensional AdS spaces, but in this paper we consider simplified action by a Weyl coupling, $\gamma$. In a general curved spacetime background, QED at 1-loop leads also to the Weyl coupling term \cite{WQED}.

One important motivation for considering such a coupling is the fact that it introduces a new parameter in the action which might affect the bounds non-trivially. One should notice that modifying the bounds in the presence of Weyl corrections might indicate some generic features of a theory which contains all possible higher derivative couplings. Indeed, in \cite{0903.2834,0811.4195} it was argued that the Weyl coupling captures the charge transport physics, significantly. They computed the correction of $\sigma$ and $D$ due to the $\gamma$, Weyl coupling. By introducing the momentum dissipation into a neutral plasma with Weyl coupling, $\sigma$ has been studied in \cite{Wu:2016jjd}. The charge response and bounds on the DC conductivity in the disordered system with higher derivative Weyl model and holographic metal-insulator transition have been studied in \cite{Fu:2017oqa} and \cite{Ling:2016dck}, respectively. Weyl holographic superconductors in the Lifshitz black hole geometry have been studied in \cite{Mansoori:2016zbp}. Holographic superconductors with momentum relaxation and Weyl coupling were studied in \cite{Ling:2016lis}. To study Weyl corrections in different setups see \cite{1510.00232,1512.07117,Mahapatra:2016dae}.

In this paper, we consider the Weyl correction in the Einstein-Maxwell action in the four-dimensional background. The momentum relaxation is introduced by a set of free axion field. First, we consider $\gamma$ as a small number and find the analytic charged black brane solution in this background as \cite{Ling:2016dck,1510.00232,1512.07117}. Next, we study bounds by computing the thermoelectric quantities in the incoherent limit. It is shown that the bound on the electric DC conductivity is violated and it should be modified in the presence of the Weyl corrections. Interestingly we find that the bound on the thermal conductivity does not change in this case. We also compute the butterfly velocity in this theory. Our results show that the Weyl coupling $\gamma$, modifies the proposed bound for $D_e$ in the incoherent limit. It has been shown also that such bounds in the inhomogeneous Sachdev-Ye-Kitaev chains can be violated \cite{Gu:2017ohj}.

This paper is organized as follows: in the next section we construct the black brane solution up to linear order in $\gamma$. We provide the reader with more details about the computations in the appendix \ref{app1} and \ref{app5}. In section 3, we compute the thermoelectric DC conductivities in the Weyl background. The technical details of the derivation of the conductivities are given in the appendix \ref{app2} and relation of holographic stress tensor and the heat current will be discussed in \ref{app4} In section 4, diffusivities in the incoherent limit have been studied. We give the details computation of the butterfly velocity in the appendix \ref{app3}. In the last section we discuss and summarize our results.

\section{Black brane solutions with Weyl correction }%
\label{setup-sec}
In this section, we consider a Weyl correction of the Maxwell field and two scalar fields in four dimensions via the following action
\begin{eqnarray}
\label{action0}
S= \int d^4 x  \sqrt{-g} \left[  R +\frac{6}{L^{2}} - \frac{1}{4} F^{\mu\nu} F_{\mu\nu}+\gamma L^{2} C_{\mu \nu \lambda \rho} F^{\mu \nu} F^{\lambda \rho}-\frac{1}{2} g^{\mu \nu} \sum_{i=1}^2 \partial_{\mu}\phi^{i}\partial_{\nu}\phi^{i}
\right],
\end{eqnarray}
where $F_{\mu\nu} =\nabla_{\mu}A_{\nu}-\nabla_{\nu}A_{\mu}$ and $C_{\mu \nu \lambda \rho}$ is the Weyl tensor \cite{0903.2834}. Because of casuality constraints and stability of the system, $\gamma$ has been limited as \cite{1010.0443}. However, involving the linear axions the bound will be modified as
\begin{equation}\label{limitg}
-\frac{1}{8}< \gamma <\frac{1}{4}.
\end{equation}
We present the derivation in the appendix \ref{app5}. The free massless scalar fields are responsible for breaking of the translational invariance and lead to momentum dissipation. Hereafter we set the AdS radius $L=1$.

From the action (\ref{action0}) one obtains the following relevant Einstein's equations,
\begin{eqnarray}\label{EinsteinEOM}
R_{\mu\nu}-{1\over2}g_{\mu\nu}R-{6}g_{\mu\nu}-T_{\mu\nu}=0.
\end{eqnarray}
\begin{eqnarray}
 \nabla_{\mu}(F^{\mu\lambda}-4\gamma  C^{\mu\nu\rho\lambda}F_{\nu\rho})=0 .
 \label{MaxwellEOM}
\end{eqnarray}
\begin{eqnarray}
 \nabla_{\mu}\nabla^{\mu} \phi^{i}=0 .
 \label{MaxwellEOM1}
\end{eqnarray}
Where $T_{\mu\nu}$ is the energy-momentum tensor,
\begin{eqnarray}
 && T_{\mu\nu}={1\over 2}\bigl(g^{\alpha\beta}F_{\mu\alpha}F_{\nu\beta}-{1\over4}g_{\mu\nu}F_{\alpha\beta}F^{\alpha\beta}\bigr)+
 {\gamma  \over 2}\bigl[g_{\mu\nu}C_{\delta\sigma\rho\lambda}F^{\delta\sigma}F^{\rho\lambda}-6 g_{\delta\mu}
 R_{\nu\sigma\rho\lambda}F^{\delta\sigma}F^{\rho\lambda} \nonumber \\
 &&-4\nabla_{\delta}\nabla_{\rho}(F^{\rho}\hspace{0.1mm}_{\mu}F^{\delta}\hspace{0.1mm}_{\nu})+2\nabla^{\sigma}
 \nabla_{\sigma}(F_{\mu}\hspace{0.1mm}^{\rho}F_{\nu\rho})+2g_{\mu\nu}\nabla_{\sigma}\nabla_{\delta}(F^{\delta}
 \hspace{0.1mm}_{\rho}F^{\sigma\rho})-4\nabla_{\delta}\nabla_{\mu}(F_{\nu\rho}F^{\delta\rho}) \nonumber \\
 && +4R_{\delta\sigma}F^{\delta}\hspace{0.1mm}_{\mu}F^{\sigma}\hspace{0.1mm}_{\nu}+8 R_{\mu\sigma}
 F^{\sigma\rho}F_{\nu\rho}-{2\over3}R_{\mu\nu}F^{\delta\sigma}F_{\delta\sigma}-{2\over3}g_{\mu\nu}\nabla^{\rho}
 \nabla_{\rho}(F^{\delta\sigma}F_{\delta\sigma}) \nonumber \\
 && +{2\over3}\nabla_{\nu}\nabla_{\mu}(F^{\delta\sigma}F_{\delta\sigma})-{4\over3}R g^{\delta\sigma}F_{\delta\mu}
 F_{\sigma\nu}\bigr]-\frac{1}{2} \sum_{i=1}^{2} \partial_{\mu}\phi^{i} \partial_{\nu} \phi^{i}+\frac{1}{4} \sum_{i=1}^{2} g_{\mu \nu}\partial_{\alpha}\phi^{i} \partial^{\alpha} \phi^{i} .
\label{EnergyMomentumTensor}
\end{eqnarray}
Now, we want to construct a black brane solution with asymptotic AdS spacetime by solving above equations. We take the following ansatz for the metric and the gauge field and the scalar fields as
\begin{eqnarray}
ds^2&=& -f(r) e^{-2\zeta(r)}  dt^2+{1\over f(r)}dr^2 +{r^2} (dx^2 + dy^2) \,,
\label{planar metric ansatz}
\end{eqnarray}
\begin{equation}
A=A_t(r)dt \,\,\,\,\ \phi^i=k x^i,\,i=1,2.
 \label{planar A ansatz}
\end{equation}
The backreaction of the gauge field on the metric makes the system difficult to solve analytically. Therefore, we will solve the system
perturbatively, up to linear order in $\gamma$ and find out the metric and the gauge field equations as \cite{Ling:2016dck,1510.00232,1512.07117,0903.2834}.
We start with the following forms for $f(r)$, $\zeta(r)$ and $A_{t}(r)$
\begin{eqnarray}\label{perturbationansatz}
f(r)&=&f^0(r)\bigl(1+\gamma \mathcal{F}(r)\bigr)\,,\nonumber\\
\zeta(r) &=& \zeta^0(r) + \gamma \zeta^1(r)\,,\\
A_{t}(r) &=& A_{t}^0(r)+ \gamma A_{t}^1(r), \ \nonumber
\end{eqnarray}
where $f^0(r)$, $\zeta^0(r)$ and $A_{t}^0(r)$ are the leading order solutions in four dimensional AdS space, with
\begin{eqnarray}
  f^{0}(r) &=& \frac{q^2}{4 r^2} + r^2 -  \frac{q^2}{4 r r_h} -  \frac{r_h^3}{r} -  \
\tfrac{1}{2} k^2  + \frac{k^2 r_h }{2 r} \,,\nonumber\\
\zeta^{0}(r) &=& 0 \,, \nonumber\\
A_{t}^{0}(r) &=& q\left( \frac{1}{r_h}- \frac{1}{r}\right).
\label{zeroth order soln}
\end{eqnarray}
Here $q$ is the charge density of the dual CFT.
Moreover, the expressions for $\mathcal{F}(r)$, $\zeta^{1}(r)$
and $A_{t}^1(r)$ can be obtained as (See Appendix \ref{app1}),
\begin{eqnarray}
\mathcal{F}(r)&=&{-1 \over f_0(r)}\left(\frac{5 k^2 q^2 r_{h}}{12 r^5}+\frac{4 k^2 q^2}{9 r^4}-\frac{k^2 q^2}{36 r r_{h}^3}-\frac{q^4}{5 r^6}+\frac{5 q^4}{24 r^5 r_{h}}-\frac{q^4}{120 r r_{h}^5}+\frac{5 q^2 r_{h}^3}{6 r^5}-\frac{4 q^2}{3 r^2}+\frac{q^2}{2 r r_{h}}\right) \,,\nonumber \\
\zeta^{1}(r)&=&  - \frac{q^2}{6 r^4} \,,\\
A_{t}^1(r)&=&\frac{k^2 q  r_{h}}{r^4}-\frac{4 k^2 q}{9 r^3}-\frac{5 k^2 q}{9  r_{h}^3}+\frac{23 q^3}{30 r^5}-\frac{q^3}{2 r^4 r_{h}}-\frac{4
   q^3}{15 r_{h}^5}-\frac{2 q  r_{h}^3}{r^4}+\frac{2 q}{ r_{h}}\nonumber.
\label{Final perturbations}
\end{eqnarray}
The Hawking temperature of the black brane is given by the following formula,
\begin{eqnarray}
T=\frac{f'(r) e^{-\zeta(r)}}{4\pi}\Big \vert_{r=r_h} =\frac{e^{\frac{\gamma  q^2}{6 r_{h}^4}} \left(-2 k^2 r_{h}^2-q^2+12 r_{h}^4\right) \left(3 r_{h}^4-2 \gamma  q^2\right)}{48 \pi  r_{h}^7},
  \label{Hawking Temp}
\end{eqnarray}
By setting $\gamma=0$, as one expects, we get back the Hawking temperature of an RN-AdS black hole.
We can also calculate the entropy of the black hole using the Wald formula \cite{9307038},
\begin{eqnarray}
S_{Wald}=-2\pi \int d^2 x \sqrt h \frac{\partial \mathcal{L}}{\partial R_{\mu\nu\rho\lambda}} \varepsilon_{\mu\nu} \varepsilon_{\rho\lambda}
\end{eqnarray}
where $\mathcal{L}$ is the Lagrangian, $\varepsilon_{\mu\nu}$ is the binormal killing vector normalised by $\varepsilon_{\mu\nu}\varepsilon^{\mu\nu}=-2$
and $h$ is the determinant of the two-sphere metric. At the leading order in $\gamma$, we arrive at
\begin{eqnarray}
&& S_{Wald}=-2 \pi  \int d^{2}x \sqrt h \bigg[\biggl(1+\frac{\gamma F_{\alpha\beta}F^{\alpha\beta}}{3}\biggr)
g^{\mu\rho}g^{\nu\lambda}\varepsilon_{\mu\nu}\varepsilon_{\rho\lambda}
+ \gamma F^{\mu\nu}F^{\rho\lambda}\varepsilon_{\mu\nu}\varepsilon_{\rho\lambda} \nonumber\\
&& \ \ \ \ \ \ \ \ \ \ - 2\gamma g^{\mu\nu}F^{\sigma\rho}F_{\rho}^{\ \lambda}\varepsilon_{\sigma\mu}\varepsilon_{\lambda\nu} \biggr]  = 4\pi r_{h}^2-\frac{8 \pi \gamma  q^2}{3 r_{h}^2} .
\label{WaldEntropy}
\end{eqnarray}
According to the gauge-string duality, we then write the charge density $q$ with respect to the chemical potential $\mu$ of the boundary field theory as follows:
\begin{equation}\label{qqqq}
q= r_{h} \mu -\gamma  \left(2 \mu  r_{h}-\frac{5 k^2 \mu }{9 r_{h}}-\frac{4 \mu ^3}{15 r_{h}}\right).
\end{equation}
\section{Thermoelectric DC conductivities}
Consider the linear response of electric current $ \mathcal{J}$ and a heat current $\mathcal{J}_{Q}$ to the small electric field $E$ and a small temperature gradient $\nabla T$. The thermo-electric conductivities are defined through the following matrix
\begin{equation}
\begin{pmatrix}
 \mathcal{J} \\
\mathcal{J}_{Q}
\end{pmatrix}
=\begin{pmatrix}
    \sigma       &  \alpha T      \\

    \bar{\alpha} T       & \bar{\kappa} T
\end{pmatrix}
\begin{pmatrix}
E \\
-\nabla T/T
\end{pmatrix}.
 \label{current matrix}
\end{equation}
In this matrix $ \sigma $ is the electric conductivity and $\alpha$, $\bar{\alpha}$ are the thermoelectric conductivities, and $\bar{\kappa}$ is the thermal conductivity.

Following the analysis in \cite{1406.4742}, one arrives at the following expressions for the full set of thermoelectric DC conductivities of the dual deformed $CFT$.
\begin{eqnarray}
&&\sigma =1+\frac{\mu ^2}{k^2}+\gamma  \left(4-\frac{4 \mu ^2}{k^2}+\frac{8 \mu ^4}{15 k^2 {r_{h}}^2}-\frac{4 k^2}{3 {r_{h}}^2}+\frac{\mu ^2}{9 {r_{h}}^2}\right)+\mathcal{O}(\gamma^2) \nonumber \\
&&\alpha  =\frac{4 \pi  \mu  {r_{h}}}{k^2}+\gamma(\frac{20 \pi  \mu }{9 {r_{h}}}-\frac{8 \pi  \mu ^3}{5 k^2 {r_{h}}}-\frac{8 \pi  \mu  {r_{h}}}{k^2})+\mathcal{O}(\gamma^2) \nonumber \\
&&\bar{\alpha} =\frac{4 \pi  \mu  {r_{h}}}{k^2}+\gamma(\frac{20 \pi  \mu }{9 {r_{h}}}-\frac{8 \pi  \mu ^3}{5 k^2 {r_{h}}}-\frac{8 \pi  \mu  {r_{h}}}{k^2})+\mathcal{O}(\gamma^2) \nonumber \\
&&\bar{\kappa}=\frac{16 \pi ^2 r_{h}^2 T}{k^2}-\gamma \frac{64 \pi ^2   \mu ^2 T}{3 k^2}+\mathcal{O}(\gamma^2).
 \label{thermoelectric and termal}
\end{eqnarray}
One finds the detailed analysis of the full DC conductivity matrix elements in the appendix \ref{app2}. Also the careful holographic renormalization of the model has been done in the appendix \ref{app4}. It is easy to check that for the case of $\gamma=0$ one can obtain the previous known results \cite{1406.4742}. In a translational invariant system, the zero-frequency conductivity has been studied in \cite{1010.0443,0811.4195}. In \cite{Donos:2017mhp} it is proved that if a model preserves the time reversal symmetry, the Onsager relation is satisfied and therefore $\alpha=\bar{\alpha}$. It is obvious that the Weyl model is invariant under the time reversal symmetry as well. Note that recently and coincidentally with this work, a similar paper has been published \cite{Chinese: 1710.07896} and the authors argue that the Weyl term breaks the time reversal invariance which is conflict with our results. We mentioned their main mistake in the appendix \ref{app2}.

Also, we define the thermal conductivity $\kappa=\bar{\kappa}-\alpha \bar{\alpha} T/ \sigma$ at zero electric current. So we have
\begin{eqnarray}
\kappa=\frac{16 \pi ^2 {r_{h}}^2 T}{k^2+\mu ^2}-\frac{16 \pi ^2 T \gamma  \left(170 k^2 \mu ^2+129 \mu
   ^4-360 \mu ^2 {r_{h}}^2\right)}{45 \left(k^2+\mu ^2\right)^2}+\mathcal{O}(\gamma^2)
   \label{kappa}
\end{eqnarray}
We express $\kappa$ in terms of horizon geometry as
\begin{eqnarray}
\kappa &=&\frac{4 \pi  f'(r)}{f''(r)}+4 \pi  \gamma f'(r)\times\nonumber\\&&\left(\frac{  -3 r^2 \zeta^{1}(r) f''(r)-16 r^2 f''(r)^2+96 r^2 f''(r)+9 r^2 f'(r) \zeta^{1}(r)'-192 r f'(r)+64 f'(r)^2}{3 r^2
   f''(r)^2}\right).\nonumber\\
\end{eqnarray}
One finds the result of \cite{Blake:2017qgd} at $\gamma=0$. Thus in the presence of Weyl corrections the formula of $\kappa$ in terms of the horizon data should be modified.

\section{Diffusivities in the incoherent limit}%
We investigate behavior of the conductivities in the incoherent limit in order to check the universal bounds proposed on the electrical conductivity and the thermal conductivity in \cite{Grozdanov:2015qia,Grozdanov:2015djs}.\\

One could consider the incoherent limit \cite{Hartnoll:2014lpa} as follows:
\begin{eqnarray}
\frac{k}{\mu } \gg 1,\quad \quad \frac{k}{T} \gg 1,\quad \quad \text{with}\quad \frac{\mu }{T}\quad  \text{finite}
   \label{incoherent limit}
\end{eqnarray}
Moreover, it is essential to clarify the relation between $r_{h}$ and $k$ in the incoherent limit. For this purpose, we expand the temperature (\ref{Hawking Temp}) as,
\begin{eqnarray}
16 \pi \left( \frac{r_{h}}{k}\right)^3 \left( \frac{T}{k}\right)&=& -2 \gamma  \delta ^2 (\frac{\mu}{k}) ^2-\frac{\gamma  }{30} (\frac{\mu}{k}) ^4-\frac{\gamma  }{9}(\frac{\mu}{k}) ^2+12
   \delta ^4-\delta ^2 (\frac{\mu}{k}) ^2-2 \delta ^2,
\end{eqnarray}
where $\delta=r_{h} / k$. Then in the incoherent limit, it is obvious that $r_{h}^{inc}= k/\sqrt{6}$.
It means that the horizon radius becomes proportional to the strength of momentum dissipation $k$.\\
Therefore in this setup, the electric, thermoelectric and thermal conductivities reduce to:
\begin{eqnarray}
&&\sigma =1- 4 \gamma +\mathcal{O}(1/k^2)   \\
&&\alpha =\bar{\alpha} =\frac{4 \pi \mu}{\sqrt{6} k}+\mathcal{O}(1/k^2)  \\
&&\kappa=\frac{8 \pi ^2T}{3}+\mathcal{O}(1/k^2) .
 \label{thermoelectric and termal incoherent}
\end{eqnarray}

As it was mentioned in the introduction, $\sigma$ in simple holographic disorder systems is bounded as \cite{Grozdanov:2015qia}
\begin{eqnarray}
\sigma \geq 1.
\end{eqnarray}
However, it is violated in the presence of Weyl corrections. It was found that the bound can be modified by considering additional couplings to the Maxwell term or with non-linear deformation \cite{Fadafan:2016gmx,Baggioli:2016oqk,Gouteraux:2016wxj,Garcia-Garcia:2016hsd,Baggioli:2016pia}.

Regarding the constraints on the $\gamma$, one may modify the bound in this model as follows:
\begin{eqnarray}
\sigma \geq \sigma_{bound}\equiv 1- 4 \gamma.\label{sb}
\end{eqnarray}

It is very interesting to study the proposed bound on the thermal conductivity in the presence of Weyl coupling. In the incoherent limits, one finds
\begin{eqnarray}
\left(\frac{\kappa}{T} \right)^{inc}=\frac{8 \pi ^2}{3}+\mathcal{O}(1/k^2)
\end{eqnarray}
It is surprising that the above relation satisfies the bound on the thermal conductivity, i.e., $\frac{\kappa}{T} \geq \mathcal{C}$ where $\mathcal{C}$ is a non zero $\mathcal{O}(1)$ number which it may rely on the different parameters and couplings of the model \cite{Grozdanov:2015djs}. It should be noted that the Weyl coupling does not affect the universal value of the bound on $\kappa/T$. \footnote{In addition to thermal conductivity bounds, we noticed that the relation:
\begin{equation}
s T \alpha- q \bar{\kappa}=0,\nonumber
\end{equation}
which was violated in \cite{Baggioli:2017ojd,Baggioli:2016pia}, does hold in our model.}

We are interested in computing $D_c$ and $D_e$ in the incoherent limit to check the universal bounds obtained in \cite{Blake:2016wvh,1604.01754}.  First of all, we need to compute the heat capacity and the charge susceptibility which are defined as follows:
\begin{equation}
C_{q}= T \left( \frac{ds}{dT} \right)_q, \quad \quad \chi= \left( \frac{\partial q}{\partial \mu } \right)_T.
\end{equation}
From Eqs. (\ref{WaldEntropy}) and (\ref{Hawking Temp}) for Wald entropy and temperature, one could derive the heat capacity formula at constant charge density as follows \cite{Mansoori:2016jer}:
\begin{eqnarray}
C_{q}= T \left( \frac{ds}{dr_{h}} \right) \left( \frac{dT}{dr_{h}} \right)^{-1}
 =\frac{128 \pi ^2 {r_{h}}^3 T}{2 k^2+\mu ^2+12 {r_{h}}^2}-\frac{64 \pi ^2 T\gamma  {r_{h}}
   \left(10 k^2 \mu ^2+3 \mu ^4+60 \mu ^2 {r_{h}}^2\right)}{15 \left(2 k^2+\mu ^2+12
   {r_{h}}^2\right)^2}
\end{eqnarray}
Moreover by making use of Eqs. (\ref{Hawking Temp}) and (\ref{qqqq}), we have
\begin{eqnarray}
\nonumber \chi &=& \bigg[\frac{\partial q}{\partial \mu} \frac{\partial T}{\partial r_{h}}- \frac{\partial T}{\partial \mu} \frac{\partial q}{\partial r_{h}}\bigg] \bigg(\frac{\partial T}{\partial r_{h}} \bigg)^{-1} =\frac{{r_{h}} \left(2 k^2+3 \mu ^2+12 {r_{h}}^2\right)}{2 k^2+\mu ^2+12 {r_{h}}^2}+\\
&&\frac{\gamma }{45 {r_{h}}} \left(2 k^2+\mu ^2+12 {r_{h}}^2\right)^{-2} \bigg[63 \mu ^4 \left(k^2+6 {r_{h}}^2\right)+4 \mu ^2 \left(k^2+6 {r_{h}}^2\right)
   \left(41 k^2+126 {r_{h}}^2\right)\nonumber \\
   &+& 20 \left(5 k^2-18 {r_{h}}^2\right) \left(k^2+6
   {r_{h}}^2\right)^2+9 \mu ^6\bigg]
\end{eqnarray}
In the incoherent limit, the heat capacity and susceptibility take the below forms, respectively.
\begin{equation}
C_{q}^{inc}=\frac{8}{3} \sqrt{\frac{2}{3}} \pi ^2 k T
\end{equation}
\begin{equation}
\chi^{inc}= \frac{k}{\sqrt{6}} \left(1+\frac{4}{3} \gamma \right)
\end{equation}

Regarding (\ref{butterfly velocity}) in the incoherent limit, the butterfly velocity also becomes:

\begin{equation}\label{vb}
{v_B^2}^{inc}=\frac{\sqrt{6} \pi  T}{k}\left(1+8\gamma\right).
\end{equation}
We present the derivation in appendix \ref{app3}.

Now we define the charge and energy diffusion as
\begin{equation}
D_{c}= \frac{\sigma}{\chi}, \quad \quad D_{e}= \frac{\kappa}{C_{q}}.
\end{equation}
Therefore, in the incoherent regime the charge and energy diffusion of our Weyl model can be written as:
\begin{equation}\label{diffusive}
\frac{D_{c} T}{v_B^2}\Big\rvert  ^{inc}= \frac{1}{\pi}(1-\frac{40}{3} \gamma), \quad \quad \frac{D_{e} T}{v_B^2} \Big\rvert  ^{inc}= \frac{1}{2\pi}\left( 1-8\gamma\right).
\end{equation}
It is obvious that both diffusivities are dependent on the Weyl coupling $\gamma$ due to the modification of the heat capacity and susceptibility and butterfly velocity. To avoid negative values for the charge diffusion, one should consider $\gamma$ in the range of $-\frac{1}{8}<\gamma<\frac{3}{40}$.  Therefore, in the presence of Weyl corrections the universal bound on the charge diffusion is violated in the incoherent limit. On the other hand, one finds that the energy diffusion is positive in the range of $-\frac{1}{8}<\gamma<\frac{1}{8}$. It means that the modified bound on $D_e$ in the presence of $\gamma$ can not be universal in the incoherent limit \cite{Blake:2016wvh,1604.01754}.

\section{Discussion}
In this paper, we studied the behavior of the thermoelectric DC conductivities in the presence of Weyl corrections with momentum dissipation in the incoherent limit. We considered the bulk Abelian action which contains the coupling of the Maxwell term and the Weyl tensor. The Weyl coupling $\gamma$, is constrained from the casualty of the dual theory. The free massless axion field is responsible for breaking the translational symmetry. Considering $\gamma$ as a small parameter, we constructed the black brane solution analytically as \cite{Ling:2016dck}. Then we computed the thermoelectric conductivities and studied them in the incoherent limit. In (\ref{sb}), it was shown that the bound on the DC conductivity, $\sigma$ is violated. Moreover, we computed the butterfly velocity and diffusion with broken translational symmetry. It was found that the Weyl coupling $\gamma$, corrects the proposed bound for the diffusivities in the incoherent limit. We also found that the bounds on the charge and energy diffusion were violated. This fact that in the incoherent limit the Weyl coupling $\gamma$, affect the $\frac{D_eT}{v_B^2}$ bound is surprising. This result is not the same as the linear axion model \cite{Blake:2016wvh,1604.01754} and Horndeski theory \cite{Baggioli:2017ojd}. One should notice that this is a highly non trivial check of the proposed bounds on the energy diffusivity and thermal conductivity which implies that it would be very important to further study them in different backgrounds.
\\\\

\textbf{Acknowledgments}
 We would like to thank J. Gauntlett, A. Lucas, R. Rodgers and R. Davison for useful discussions. Especially thank A. O'Bannon for reading the manuscript and comments. We also thank V. Jahnke for discussions on the butterfly velocity. KBF wishes to thank the University of Southampton for hospitality during the course of this work. We would like to thank the referee for his/her instructive comments.

\appendix
\section{Calculating the black brane solution}\label{app1}%
The functions $\mathcal{F}(r)$, $\zeta^{1}(r)$ and $A_{t}^{1}(r)$ are obtained by solving
Eqs.(\ref{EinsteinEOM}) and (\ref{MaxwellEOM}) up to the linear order in $\gamma$,
\begin{eqnarray}
\mathcal{F}(r)&=& {1 \over f_0(r)} \bigl(\frac{k_2}{r}-\frac{5 k^2 q^2 r_{h}}{12 r^5}+\frac{4 k^2 q^2}{9 r^4}-\frac{q^4}{5 r^6}+\frac{5 q^4}{24 r^5 r_{h}}+\frac{5 q^2 r_{h}^3}{6 r^5}-\frac{4 q^2}{3 r^2}\bigr) \,,\nonumber\\
\zeta^1(r)&=&  - \frac{q^2}{6 r^4}+k_{1} \,,\\
A_{t}^1(r)&=& -\frac{k_3}{r}+k_4-\frac{k^2 q r_{h}}{2 r^4}+\frac{2 k^2
   q}{9 r^3}-\frac{13 q^3}{30 r^5}+\frac{q^3}{4 r^4
   r_{h}}+\frac{q r_{h}^3}{r^4} .\ \nonumber
\label{perturbations}
\end{eqnarray}
where $k_1$, $k_2$, $k_3$, and $k_4$ are dimensionless integration constants and $r_h$ in above relations indicates the position of the event horizon of the black hole.
Now, we determine those constants by imposing several constraints on the above functions as \cite{1510.00232,1512.07117,0903.2834}.
To evaluate $k_1$, we consider the asymptotic behaviour of the black hole metric,
\begin{eqnarray}
ds^2 \vert_{r\rightarrow \infty}=- (f e^{-2\chi})_\infty dt^2+r^2 (dx^2+ dy^2) .
\label{CFT metric}
\end{eqnarray}
where $(f e^{-2\zeta})_\infty=\lim_{r\to\infty}f(r) e^{-2\zeta(r)}$. It corresponds to the geometry of the dual CFT at the boundary. The speed of light in the dual theory should be unity, which requires that $(f e^{-2\zeta})_\infty={r^2 }$. Then one finds that $k_1=0$.
We also count the value of $k_3$ by requiring that the charge density $q$ remains fixed. Moreover, one could write the Maxwell equation,
eq.(\ref{MaxwellEOM}) in the form $\nabla_{\mu}X^{\mu\lambda}=0$, where, $X^{\mu\lambda}$ is an antisymmetric tensor. Therefore, one can define the
dual of $(*X)_{x y}$  which is a constant and appropriate to consider this constant as the fixed charge density $q$. Because the quantity $(*X)_{x y}$ does not depend on $r$, one may define
\begin{eqnarray}
\lim_{r\rightarrow\infty}\left(*X\right)_{x y}=q .
\label{constraint1a}
\end{eqnarray}
The left hand side of above equation in the asymptotic limit can be calculated as,
\begin{eqnarray}
\lim_{r\rightarrow\infty}\left(*X\right)_{x y}&=&\lim_{r\rightarrow\infty}\bigl[r^2  e^{\zeta(r)}
\left(F_{r t}-8\gamma L^2 C_{r t}\hspace{0.1mm}^{r t}F_{r t}\right)\bigr] \nonumber\\
&=&\bigl(1 + \gamma k_3 \bigr)q .
\label{constraint1b}
\end{eqnarray}
Now comparing the (\ref{constraint1a}) with (\ref{constraint1b}), we arrive at $k_3=0$.
On the other hand, in order to determine $k_2$, one needs to impose the condition, $f_0(r) \mathcal{F}(r) |_{r=r_h}=0$.  Thus we obtain $k_2$ as,
\begin{eqnarray}
 k_2=-\frac{k^2 q^2}{36 r_{h}^3}-\frac{q^4}{120 r_{h}^5}+\frac{q^2}{2 r_{h}}.
 \label{constraint2}
\end{eqnarray}
Since the $A_t$ vanishes at the horizon, we can count up the last constant $k_4$. Using this condition, one concludes that $A_{t}^1(r_{h})=0$ and the constant $k_4$ is given by,
\begin{eqnarray}
  k_4=-\frac{5 k^2 q}{9  r_{h}^3}-\frac{4 q^3}{15  r_{h}^5}+\frac{2 q}{ r_{h}} .
 \label{constraint3}
\end{eqnarray}
\section{Thermoelectric transports calculations }\label{app2}
In order to compute the conductivities, we follow the procedure established in \cite{1406.4742}. Therefore, we consider the following perturbations,
\begin{eqnarray}
 g_{tx} &\rightarrow& t\delta h(r) + \delta g_{tx} (r), \nonumber \\
 g_{rx} &\rightarrow & r^{2} \delta  g_{rx} (r), \nonumber \\
 A_{x} &\rightarrow &  t\delta a(r)  + \delta A_{x} (r) \nonumber\\
 \phi_1  &\rightarrow &  k\,x+ \delta \phi_1 (r).
  \label{perturbation of metric and maxwel}
\end{eqnarray}
By substituting these perturbations in (\ref{EinsteinEOM}), one finds the  $xx$ and $rx$ components of Einstein equations \footnote{One should notice that our results (\ref{xx}) and (\ref{rx}) disagree with that in \cite{Chinese: 1710.07896}. We checked that there are some missing terms in that paper.}. The $xx$ Einstein equation is given by
\begin{eqnarray}\label{xx}
&-& \frac{3}{2} r^2 f' \zeta '+\frac{1}{2} r^2 f''+r f'-r^2 f \zeta ''+r^2 f \zeta '^2-r f \zeta '-\frac{1}{4} r^2 e^{2 \zeta }
A_t'^2-3 r^2-\frac{1}{3} \gamma  r^2 e^{2 \zeta } f'' A_t'^2\nonumber\\&+&\frac{5}{3} \gamma  r^2 e^{2 \zeta } f' A_t'^2 \zeta '+\frac{2}{3} \gamma
r^2 e^{2 \zeta } f' A_t' A_t''+\frac{4}{3} \gamma  r e^{2 \zeta } f' A_t'^2+
\frac{2}{3} \gamma  r^2 f
e^{2 \zeta}A_t''^2+\frac{4}{3} \gamma  r^2 f e^{2 \zeta} A_t'^2 \zeta ''\nonumber\\&+&\frac{2}{3} \gamma  r^2 f
A_t''' e^{2 \zeta } A_t'+2 \gamma  r^2 f e^{2 \zeta } A_t' A_t'' \zeta '+
\frac{4}{3} \gamma  r
f e^{2 \zeta } A_t'^2 \zeta '+\frac{8}{3} \gamma  r f e^{2 \zeta } A_t' A_t''=0
\end{eqnarray}
Also the  $rx$ Einstein equation is given by
\begin{eqnarray}\label{rx}
&&\frac{1}{2} r^2 \delta  g_{rx} f''-\frac{3}{2} r^2 \delta  g_{rx} f' \zeta '+r \delta  g_{rx} f'-r^2 f \delta  g_{rx} \zeta ''+r^2 f \delta  g_{rx} \zeta '^2-\frac{\delta a e^{2 \zeta} A'}{2 f}-\nonumber\\&&
r f\delta  g_{rx} \zeta '-\frac{e^{2 \zeta } \delta h'}{2 f}+\frac{\delta h e^{2 \zeta}}{r f}+\frac{1}{2} k^2 \delta  g_{rx}-\frac{1}{2} k \delta \phi_ 1'-\frac{1}{4} r^2\delta  g_{rx} e^{2 \zeta } A'^2-3 r^2 \delta  g_{rx}-\nonumber\\&&
\frac{1}{3} \gamma  r^2 \delta g_{rx} e^{2 \zeta } f'' A'^2-\frac{2 \gamma  \delta a e^{2 \zeta } f''
	A'}{3 f}+\frac{5}{3} \gamma  r^2 \delta g_{rx} e^{2 \zeta } f' A'^2 \zeta '+\frac{2}{3} \gamma  r^2
\delta g_{rx} e^{2 \zeta } f' A' A''+\nonumber\\&&\frac{\gamma  e^{2 \zeta } f' A' \delta
		a'}{f}+\frac{2 \gamma  \delta a e^{2 \zeta } f' A' \zeta '}{f}+\frac{4 \gamma  \delta a e^{2 \zeta
		} f' A'}{3 r f}+\frac{4}{3} \gamma  r \delta g_{rx} e^{2 \zeta } f' A'^2+\frac{2}{3} \gamma  r^2 f
\delta g_{rx} e^{2 \zeta } A''^2+\nonumber\\&&
\frac{4}{3} \gamma  r^2 f \delta g_{rx} e^{2 \zeta } A'^2 \zeta
''+
\frac{2}{3} \gamma  r^2 f A^{'''}  \delta g_{rx} e^{2 \zeta } A'+2 \gamma  r^2 f  \delta g_{rx}
e^{2 \zeta } A' A'' \zeta '+\frac{4}{3} \gamma  r f  \delta g_{rx} e^{2 \zeta } A'^2 \zeta
'+\nonumber\\&&
\frac{4 \gamma  e^{4 \zeta } A'^2 \delta h'}{3 f}-\frac{8 \gamma  \delta h e^{4 \zeta }
	A'^2}{3 r f}+\frac{8}{3} \gamma  r f  \delta g_{rx} e^{2 \zeta } A' A''-\frac{4 \gamma 
	\delta a e^{2 \zeta } A'}{3 r^2}+\gamma  e^{2 \zeta } A'' \delta a'+\gamma  e^{2 \zeta }
A' \delta a''+\nonumber\\&&
\frac{4}{3} \gamma  \delta a e^{2 \zeta } A' \zeta ''-\frac{4}{3} \gamma 
\delta a e^{2 \zeta } A' \zeta '^2-\frac{4 \gamma  \delta a e^{2 \zeta } A' \zeta '}{3 r}=0
\end{eqnarray}
By resolving $f''$ from (\ref{xx}), the above equation simplified as
\begin{eqnarray}
&&\frac{\gamma  e^{2 \zeta } f' A' \delta a'}{f}+\frac{8 \gamma  \delta a e^{2 \zeta } f'
	A'}{3 r f}-\frac{\gamma  \delta a e^{4 \zeta } A'^3}{3 f}-\frac{4 \gamma  \delta a e^{2 \zeta
		} A'}{f}+\frac{4 \gamma  e^{4 \zeta } A'^2 \delta h'}{3 f}-\frac{8 \gamma  \delta h e^{4
		\zeta } A'^2}{3 r f}-\nonumber\\&&
\frac{\delta a e^{2 \zeta } A'}{2 f}-\frac{e^{2 \zeta } \delta h'}{2
	f}+\frac{\delta h e^{2 \zeta }}{r f}+\frac{1}{2} k^2 \delta g_{rx}-\frac{1}{2} k \delta \phi_1'-\frac{4 \gamma
	\delta a e^{2 \zeta } A'}{3 r^2}+\gamma  e^{2 \zeta } A'' \delta a'+\gamma  e^{2 \zeta }
A' \delta a''-\nonumber\\&&
\frac{8 \gamma  \delta a e^{2 \zeta } A' \zeta '}{3 r}=0
\end{eqnarray}
Regarding the perturbation in (\ref{perturbationansatz}), one gets the $rx$ Einstein equation (\ref{rx}) up to the linear order in $\gamma$ as follows:
\begin{equation}
\delta \phi'_{1} - k \delta  g_{rx}+ \frac{r^2}{k f^0 } \left[ \frac{\delta h }{r^2} \right]^{'}+ \frac{\delta a {A_{t}^{0}}'}{k f^0 } +\gamma K_{0}=0,
 \label{Einstain rx}
\end{equation}
where $K_{0}$ is a function as follows:
\begin{eqnarray}
K_{0}&&=\frac{\delta a {A_{t}^{0}}'}{k f^{0}}\left[ 8 +\frac{8 f^0}{3 r^2} -\mathcal{F} +2 \zeta^{1} + \frac{2}{3}{({A_{t}^{0}}')}^2-\frac{16}{3 r} {f^0}' \right]+ \frac{\delta a {A_{t}^{1}}'}{k f^{0}} \nonumber\\
&&+\frac{1}{k f^{0}} \left[  \left[ \frac{\delta h}{r^2} \right]^{'} \left[ -\frac{8 r^2}{3}  {A_{t}^{0}}'^2- r^2 \mathcal{F}+2r^2 \zeta^{1}  \right]-2\left[ \delta a' {A_{t}^{0}}' f^0 \right]^{'} \right].
\label{K(r)}
\end{eqnarray}
Now we can solve the $\delta g_{rx}$ as,
\begin{equation}
\delta  g_{rx}=\frac{\delta \phi'_{1}}{k } + \frac{r^2}{k^2 f^0 } \left[ \frac{\delta h}{r^2} \right]^{'}+ \frac{\delta a {A_{t}^{0}}'}{k^2 f^0 } +\frac{\gamma}{k } K_{0}.
 \label{Grx EOM}
\end{equation}

In addition, from Maxwell equation (\ref{MaxwellEOM}), one can easily derive a radially conserved current called electric current $\mathcal{J}$ as
\begin{equation}
\mathcal{J}=-\sqrt{-g} (F^{rx}-4\gamma C^{r x \rho\lambda}F_{\rho\lambda})
    \label{electric current}
\end{equation}
in which for our model, it yields:
\begin{eqnarray}
\mathcal{J}=&&-\frac{4}{3} \gamma  \delta g_{tx} {A^{0}}' {f^{0}}''-\frac{4}{3} \gamma  t \delta h {A^{0}}'
   {f^{0}}''+\frac{8 \gamma  \delta g_{tx} {A^{0}}' {f^{0}}'}{3 r}+\frac{8 \gamma  t \delta h
   {A^{0}}' {f^{0}}'}{3 r}+\frac{4 \gamma  {f^{0}} \delta g_{tx} {A^{0}}'}{3 r^2}\nonumber\\
&& +\frac{4 \gamma  t{f^{0}} \delta h {A^{0}}'}{3 r^2}+2 \gamma  {f^{0}} {A^{0}}' \delta g_{tx}''-\frac{4 \gamma
   {f^{0}} {A^{0}}' \delta g_{tx}'}{r}+2 \gamma  t {f^{0}} {A^{0}}' \delta h''-\frac{4 \gamma  t
   {f^{0}} {A^{0}}' \delta h'}{r}\nonumber\\
&& -\gamma  \delta g_{tx} \zeta^{1} {A^{0}}'-\delta g_{tx} {A^{0}}'-\gamma  t \delta h \zeta^{1} {A^{0}}'-t \delta h {A^{0}}'-\gamma
   \delta g_{tx} {A^{1}}'-\gamma  {\mathcal{F}} {\delta A_{x}}'-\gamma  t {\mathcal{F}} \delta a' \nonumber\\
&& -\gamma  t \delta h {A^{1}}'+\frac{2}{3} \gamma  {f^{0}} {f^{0}}''
   {\delta A_{x}}'+\frac{2}{3} \gamma  t {f^{0}} {f^{0}}'' \delta a'-\frac{4 \gamma  {f^{0}} {f^{0}}'
   {\delta A_{x}}'}{3 r}-\frac{4 \gamma  t {f^{0}} {f^{0}}' \delta a'}{3 r} \nonumber\\
&&+\frac{4 \gamma  {f^{0}}^2 {\delta A_{x}}'}{3 r^2}+\frac{4 \gamma  t {f^{0}}^2 \delta a'}{3 r^2}+\gamma  {f^{0}} \zeta^{1}
   {\delta A_{x}}'-{f^{0}} {\delta A_{x}}'+\gamma  t {f^{0}} \zeta^{1} \delta a'-t {f^{0}}
   \delta a'.
    \label{electric current 1}
\end{eqnarray}
For a general class of gravity theories, the Noether theorem can also be used to derive a general formula for the radially conserved heat current in AdS planar black holes with certain transverse and traceless perturbations,
\begin{equation}
\mathcal{G}_{gravity}^{\mu\nu}= 2 \frac{\partial \mathcal{L}}{\partial R_{\mu \nu \rho \sigma}} \nabla_{\rho} \zeta_{\sigma} -4\zeta_{\sigma} \nabla_{\rho}\frac{\partial \mathcal{L}}{\partial R_{\mu \nu \rho \sigma}},
    \label{heat current}
\end{equation}
where $\mathcal{L}$ is the Lagrangian and $\zeta_{\sigma}$ is an infinitesimal diffeormorphism $\zeta_{\sigma}=\delta x_{\sigma}$ \cite{9403028,0911.5004,1708.02329,1306.2138}. For the Einstein-Hilbert term, for instance, we have compacted form $\mathcal{G}_{gravity}^{\mu\nu}=2 \nabla^{\mu} \zeta^{\nu}$. Then for an arbitrary Killing vector $\zeta^{\mu}$, we can define a two-form $\mathcal{H}$ by:
\begin{equation}\label{Hcurrent}
\mathcal{H}^{\mu\nu}=\mathcal{G}_{gravity}^{\mu\nu}+\mathcal{G_{A}}^{\mu\nu}
\end{equation}
 Note that the second contribution is associated with the minimally-coupled Maxwell field which for our model,
 \begin{equation}
 \mathcal{G_{A}}^{\mu\nu}=\zeta^{\sigma} A_{\sigma} (F^{\mu \nu}-4\gamma C^{\mu\nu \rho\lambda}F_{\rho\lambda})
\end{equation}
 For a timelike Killing vector $\zeta=\partial_{t}$, we consider the $x$ component of (\ref{Hcurrent}) to deduce that $\partial_{r} (\sqrt{-g} \mathcal{H}^{rx})=0$ and thus $\sqrt{-g} \mathcal{H}^{rx}$ is a constant. It is worth noting that for the electric current term $\mathcal{J}$, we arrive at the same result as (\ref{electric current}). Finally, the heat current is defined as,
 \begin{equation}\label{Heat current}
 \mathcal{J_{Q}}=\sqrt{-g} \mathcal{H}^{rx}
 \end{equation}
Substituting now the perturbations (\ref{perturbation of metric and maxwel}) in equation (\ref{Heat current}), we obtain,
\begin{eqnarray}
\mathcal{J}_{Q}=&& 2 \gamma  {f^{0}}^2 {A^{0}}'' {\delta A_{x}}'+2 \gamma  t {f^{0}}^2 {A^{0}}'' {\delta a}'+2 \gamma  {f^{0}} {A^{0}}' {f^{0}}'
   {\delta A_{x}}'+\frac{2}{3} \gamma  {\delta g_{tx}} {A^{0}}'^2 {f^{0}}'+2 \gamma  t {f^{0}} {A^{0}}' {f^{0}}' {\delta a}' \nonumber\\
  && +\frac{2}{3} \gamma  t {\delta h} {A^{0}}'^2 {f^{0}}'+2 \gamma  {f^{0}}^2 {A^{0}}' {\delta A_{x}}''+\frac{4 \gamma  {f^{0}}^2
   {A^{0}}' {\delta A_{x}}'}{r}+\frac{4}{3} \gamma  {f^{0}} {A^{0}}'^2 {\delta g_{tx}}'+\frac{4 \gamma  {f^{0}} {\delta g_{tx}}
   {A^{0}}'^2}{r}\nonumber\\
   &&+2 \gamma  t {f^{0}}^2 {A^{0}}' {\delta a}''+\frac{4 \gamma  t {f^{0}}^2 {A^{0}}' {\delta a}'}{r}+\frac{4}{3} \gamma
   t {f^{0}} {A^{0}}'^2 {\delta h}'+\frac{4 \gamma  t {f^{0}} {\delta h} {A^{0}}'^2}{r}+4 \gamma  {f^{0}} {\delta g_{tx}}
   {A^{0}}' {A^{0}}''\nonumber\\
   &&+4 \gamma  t {f^{0}} {\delta h} {A^{0}}' {A^{0}}''+\gamma  {\delta g_{tx}} {\zeta^{1}}
   {f^{0}}'-{\delta g_{tx}} {f^{0}}'+\gamma  t {\delta h} {\zeta^{1}} {f^{0}}'-t {\delta h} {f^{0}}'-\gamma  {f^{0}}
   {\zeta^{1}} {\delta g_{tx}}'\nonumber\\
   &&+2 \gamma  {f^{0}} {\delta g_{tx}} {\zeta^{1}}'+{f^{0}} {\delta g_{tx}}'-\gamma  t {f^{0}}
   {\zeta^{1}} {\delta h}'+2 \gamma  t {f^{0}} {\delta h} {\zeta^{1}}'+t {f^{0}} {\delta h}'-\gamma  {\delta g_{tx}}
   {\mathcal{F}}'-\gamma  t {\delta h} {\mathcal{F}}'\nonumber\\
   &&+\gamma  {\mathcal{F}} {\delta g_{tx}}'+\gamma  t {\mathcal{F}} {\delta h}'-A_{t} \mathcal{J}
    \label{heat current 1}
\end{eqnarray}
Next, we assume ${\delta a}(r)=-E+C A_{t}(r)$ and ${\delta h}(r)=-C f(r) e^{-2 \zeta(r)}$ in order to cancel the terms depend on $t$ in $\mathcal{J}$ and $\mathcal{J}_{Q}$ in equations (\ref{electric current 1}) and (\ref{heat current 1}). One also finds the near horizon behavior of $\delta g_{rx}$ as 
\begin{eqnarray}
\delta g_{rx}&\approx & \frac{E}{f^{0}} \Big[\frac{16 \gamma  {A'}^{0}_{t} {f'}^{0}}{3 k^2 r}-\frac{2 \gamma  {A'}^{0}_{t}(r)^3}{3 k^2}-\frac{8 \gamma 
	{A'}^{0}_{t}(r)}{k^2}-\frac{{A'}^{0}_{t}(r)}{k^2}-\frac{\gamma  {A'}^{1}_{t}}{k^2}\Big]\\
\nonumber &&+\frac{C}{f^{0}} \Big[\frac{2 \gamma  {A'}^{0}_{t}(r)^2 {f'}^{0}(r)}{3 k^2}+\frac{2 \gamma  \zeta^{1}(r) {f'}^{0}(r)}{k^2}-\frac{\gamma  \mathcal{F}(r)
	{f'}^{0}(r)}{k^2}-\frac{{f'}^{0}(r)}{k^2}\Big]
\end{eqnarray}
Now we consider the asymptotic behavior near the horizon $r = r_h$. Since we consider the boundary condition at the future horizon, we will use ingoing Eddington-Finklestein
coordinates $(v, r)$ defined as $v = t + \int \frac{dr}{e^{- \zeta(r)} f(r)}$ here.
First, the gauge field should be regular at the future horizon, so from equation (\ref{perturbation of metric and maxwel}) we conclude that $\delta A_x$ should satisfy
\begin{equation}
\delta A_{x}=-E \int \frac{dr}{e^{- \zeta(r)} f(r)}
\end{equation}
near horizon $r = r_h$. In addition, we can find that $\delta g_{rx}\sim \frac{1}{r-r_h}$ is divergence as $r \rightarrow r_h$. Therefore, in order to obtain the singular part of the metric in the ingoing Eddington-Finklestein coordinates we should require the metric perturbation behaves as
\begin{equation}
\delta g_{tx}= r^2e^{- \zeta(r)} f(r) \delta g_{rx}-C e^{- 2 \zeta(r)} f(r) \int \frac{dr}{e^{- \zeta(r)} f(r)}.
\end{equation}
We can now compute the electric and heat currents at the horizon. Therefore, the thermoelectric conductivities can be defined as:
\begin{eqnarray}
\sigma= \frac{\partial \mathcal{J}(r_{h})}{\partial E},\quad \alpha=\frac{1}{T} \frac{\partial \mathcal{J}(r_{h})}{\partial C},\quad \bar{\alpha}=\frac{1}{T} \frac{\partial \mathcal{J}_{Q}(r_{h})}{\partial E}, \quad \bar{\kappa}=\frac{1}{T} \frac{\partial \mathcal{J}_{Q}(r_{h})}{\partial C}.
\end{eqnarray}
where all the quantities have to be computed at the horizon $r=r_{h}$. One obtains following equations
\begin{eqnarray}
&&\sigma=1+\frac{r^2 A_{t}'(r)^2}{k^2}- \frac{2\gamma}{3k^2r} \bigg[ -2 r^3 A_{t}'(r)^2 f''(r)+12 r^2 A_{t}'(r)^2 f'(r)-\nonumber\\
&&\,\,\,\,\,\,\,\,3 r^3 \zeta^1(r) A_{t}'(r)^2-r^3 A_{t}'(r)^4-12 r^3 A_{t}'(r)^2+k^2 r f''(r)-2 k^2 f'(r))\bigg], \nonumber \\
&&\alpha=\frac{4 \pi  r^2 A_{t}'(r)}{k^2}+\frac{16 \pi  \gamma  r^2 A_{t}'(r) f''(r)}{3 k^2}-\frac{32 \pi  \gamma  r A_{t}'(r) f'(r)}{3 k^2}+\frac{4 \pi  \gamma  r^2 \zeta^1(r) A_{t}'(r)}{k^2}-\frac{8 \pi  \gamma
   r^2 {A'}_{t}(r)^3}{3 k^2}, \nonumber \\
&&\bar{\alpha}=\frac{4 \pi  r^2 A_{t}'(r)}{k^2}-\frac{64 \pi  \gamma  r A_{t}'(r) f'(r)}{3 k^2}+\frac{4 \pi  \gamma  r^2 \zeta^1(r) A_{t}'(r)}{k^2}+\frac{32 \pi  \gamma  r^2 A_{t}'(r)}{k^2},\nonumber\\
&&\bar{\kappa}=\frac{4\pi r^2 f'(r)}{k^2}+\frac{4 \gamma  \left(-4 \pi  r^2 A_{t}'(r)^2 f'(r)^2-3 \pi  r^2 \zeta^1(r) f'(r)^2\right)}{3 k^2 f'(r)}.
 \label{thermoelectric and thermal}
\end{eqnarray}
As it seems the terms contain $\gamma$ in  $\alpha$  and $\bar{\alpha}$ are not equal, but as we show below they are exactly equal. We start with the $r$ component of the Maxwell equation which is,
 \begin{eqnarray}\label{MMEQ}
 &&8\gamma r e^{\zeta } A_{t}' f'\zeta' + \frac {8} {3}\gamma r^2 f e^{\zeta}\zeta''' A_{t}' - \frac {8} {3} \gamma r^2 f e^{\zeta } A_{t}'\zeta'^3 - \frac{8} {3}\gamma r^2 f e^{\zeta } A_{t}'\zeta'\zeta'' + \frac {8} {3}\gamma r f e^{\zeta } A_{t}'\zeta'' \nonumber \\
&&-\frac {4} {3}\gamma r^2 e^{\zeta } A_{t}'' f'' + 4\gamma r^2 e^{\zeta } A_{t}'' f'\zeta' + \frac {8} {3}\gamma r e^{\zeta } A_{t}'' f' + \frac {8} {3}\gamma r^2 f e^{\zeta } A_{t}''\zeta'' - \frac {8}{3}\gamma r^2 f e^{\zeta } A_{t}''\zeta'^2  \nonumber \\
&&-\frac {8} {3}\gamma f e^{\zeta } A_{t}'' - \frac {4} {3}\gamma r^2 f''' e^{\zeta } A_{t}' + \frac {8} {3}\gamma r^2 e^{\zeta } A_{t}' f''\zeta' + \frac {20}{3}\gamma r^2 e^{\zeta } A_{t}' f'\zeta'' + \frac {4}{3}\gamma r^2 e^{\zeta } A_{t}' f' \zeta'^2 -  r^2 e^{\zeta } A_{t}''\nonumber  \\
&&-8\gamma r f e^{\zeta } A_{t}'\zeta'^2 - \frac {16} {3}\gamma f e^{\zeta } A_{t}'\zeta' - r^2 e^{\zeta } A_{t}'\zeta' - 2 r e^{\zeta } A_{t}' - \frac {8} {3}\gamma r f e^{\zeta } A_{t}'' \zeta'  = 0.
 \end{eqnarray}
In addition the $(r,r)$ component of the Einstein equation is,
 \begin{eqnarray}\label{EEEQ}
&-&\frac{8 \gamma e^{2 \zeta } f' {A'}_{t}^2 \zeta '}{3
f}+\frac{2 \gamma e^{2 \zeta } f'' {A'}_{t}^2}{3 f}-\frac{2 \gamma e^{2 \zeta } f' {A'}_{t}^2}{r f}-\frac{2 \gamma e^{2 \zeta } f' {A'}_{t}
   {A''}_{t}}{3 f}
   +\frac{f'}{r f}+\frac{k^2}{2 r^2 f}+\frac{e^{2 \zeta } {A'}_{t}^2}{4 f}\nonumber \\&-&\frac{3}{f} +\frac{8 \gamma
    e^{2 \zeta } {A'}_{t}^2}{3 r^2}+\frac{1}{r^2}-\frac{4}{3} \gamma e^{2 \zeta } {A'}_{t}^2 \zeta ''+\frac{8}{3}
   \gamma e^{2 \zeta } {A'}_{t}^2 \zeta '^2+\frac{4 \gamma e^{2 \zeta } {A'}_{t}^2 \zeta '}{r} + \frac{4}{3}
   \gamma e^{2 \zeta } {A'}_{t} {A''}_{t} \zeta ' \nonumber \\ &+&\frac{4 \gamma e^{2 \zeta } {A'}_{t} {A''}_{t}}{3
   r}-\frac{2 \zeta '}{r}=0
\end{eqnarray}
Notice that one can ignore the terms which are non linear order in $\gamma$ like $\gamma e^{2 \zeta }, e^{2 \zeta } \zeta ', e^{2 \zeta } \zeta '' $. By solving the $A_{t}''(r)$ in both of the equations (\ref{MMEQ}) and (\ref{EEEQ}) and comparing them to each other, one can solve the $A_{t}'(r)$ in terms of the metric components and $k$. 
Next by substituting the $A_{t}'(r)$, $A_{t}''(r)$, and $A_{t}'''(r)$ in $xx$ component of the Einstein equation (\ref{xx}), we can also solve $k$ in terms of the metric components.
Then by replacing $A'(r)$ and $k$ in those terms which do not seem to be equal in $\alpha$ and $\bar{\alpha}$ relations and by evaluating at the horizon one obtains
\begin{eqnarray}\label{alpha=alphabar}
\alpha=\bar{\alpha}=\frac{4 \pi  r^2 A_{t}'(r)}{k^2}-\frac{4 \sqrt{2} \pi  \gamma  \left(3 r (\zeta^1(r)+8)-16 f'(r)\right)\sqrt{r f''(r)-6 r^2 +2 r f'(r)}}{
 3 r^2 f''(r)-36 r^2 +12 r f'(r)}.
\end{eqnarray}
Note that we have used the assumption of $\delta \phi_1$ is regular at the horizon.
\section{Butterfly velocity calculation}\label{app3}
In a quantum system, the butterfly effect can be characterized by the out-of-time-order corrector (OTOC). After the scrambling time $t_{*}$, the exponential deviation of OTOC takes the form,
\begin{eqnarray}
\frac{\langle V_{x}(0)W_{y}(t)V_{x}(0)W_{y}(t)\rangle_{\beta}}{\langle V_{x}(0)V_{x}(0)\rangle_{\beta} \langle W_{y}(t)W_{y}(t)\rangle_{\beta}}\sim 1-e^{\lambda_{L}\big(t-t_{*}-\frac{|x-y|}{v_{B}}\big)},
\end{eqnarray}
 where $V$ and $W$ are two generic Hermitian operators. Furthermore, $ \lambda_{L} $ is the Lyapunov exponent and $ v_{B} $ is the butterfly velocity. The Lyapunov exponent is, $ \lambda_{L}=\frac{2\pi}{\beta} $, where $ \beta $ is inverse of Hawking temperature. And also the butterfly velocity should be identified by the velocity of shock wave by which the perturbation spreads in the space \cite{1306.0622,Roberts:2014isa,1412.6087}. Moreover, the butterfly effect is obtained in D-dimensional gravitational theory containing higher order derivatives and Massive gravities \cite{1610.02890,1705.05235,1707.00509}. Now we study a shock wave solution of our model when the above black hole
solution is perturbed by injecting a small amount of energy. To proceed, it is useful to re-write the black brane solution in the Kruskal coordinates:
\begin{eqnarray}\label{d}
u\,v\,=\,-\,e^{\sqrt{f'(r_h)\,h'(r_h)}\,r_*}, \ \ \ u/v\,=\,-\,e^{-\,\sqrt{f'(r_h)\,h'(r_h)}\,t},
\end{eqnarray}
where $dr_*=\frac{dr}{\sqrt{f(r)h(r)}}$ and $h(r)=f(r) e^{-2\zeta(r)}$. By making use of this coordinate system, the metric and the gauge filed can be recast into the following form \footnote{In the first version, we had a mistake in defining $A$ in the Kruskal coordinate.}
\begin{eqnarray}\label{met1}
 \nonumber ds^2\,&=&\,2\,A(uv)\,du\,dv\,+\,B(uv)\,dx^i\,dx^i,\\
A&=&A_{\mu} dx^{\mu}= -\Phi(uv) v du+\Phi(uv) u dv
\end{eqnarray}
 where functions appearing in the metric are related by the following relations:
\begin{eqnarray}\label{e}
A(uv)\,=\,\,\frac{2}{u\,v}\,\frac{h(r)}{f'(r_h)\,h'(r_h)}, \ \ B(uv)\,=\,r^2, \ \ \Phi(uv)\,=\,\frac{1}{uv} \frac{A_{t} (r)}{\sqrt{f'(r_{h}) h'(r_{h})}}
\end{eqnarray}
for which the horizon location $r=r_h$ is mapped into $uv=0$. We perturb it by releasing a particle from the boundary at $x=0$. For the late time $t_{w}>\beta$
the localized stress tensor resulted from this particle is given by
\begin{equation}
\delta T_{uu}^{shock}\,=\,E\,e^{2\,\pi\,t_w/\beta}\,\delta(u)\,\delta(x)
\end{equation}
where $\beta=1/T$ and $E$ denotes the asymptotic energy of the particle. The back reaction of this pulse of energy in the left side of the geometry is obtained when there is a shift $v\rightarrow v\,+\,h(x,t_w)$, where $h(x,t_w)$ is a function that can be determined from Einstein's equations. So we have,
\begin{equation}\label{met2}
ds^2\,=\,2\,A(uv)\,du\,dv\,+\,B(uv)\,dx^i\,dx^i\,-\,2\,A(uv)\,h(x,t_w)\,\delta(u)\,du^2
\end{equation}
Now let us find the function $h(x,t_w)$ such that the above ansatz satisfies the Einstein's equations,
\begin{equation}
G_{\mu\nu}+\delta G_{\mu\nu}=T_{\mu\nu}+\delta T_{\mu\nu}+\delta T_{\mu\nu}^{shock}
\end{equation}
Assuming the Einstein's equations (\ref{EinsteinEOM}) are satisfied in background metric (\ref{met1}), one can derive a second order differential equation for $h(x,t_w)$ near the horizon at the leading order of the perturbation by plugging the perturbed ansatz metric (\ref{met2}) in Einstein equation as
\begin{eqnarray}\label{c4}
\left[\partial_i^2-m^2\right]h(x^i,t_w)=\frac{3 A(0)^3 B(0)E e^{2\pi t_w/\beta}\delta(x)}{
   16 \gamma
    A(0)^2 \Phi (0)^2+3
   A(0)^4}
\end{eqnarray}
where the effective mass reads:
\begin{eqnarray}\label{c5}
m^2=\frac{8 \gamma  \Phi (0) \left(-2 B(0) \Phi (0)
   A'(0)+A(0) \Phi (0) B'(0)+4 A(0) B(0) \Phi '(0)\right)-3
   A(0)^3 B'(0)}{16 \gamma  A(0)^2 \Phi (0)^2+3
   A(0)^4}
\end{eqnarray}
 At large distances $|x|$, the solution takes the form:
\begin{equation}\label{reH}
h(x,t_w)\,\sim\,\frac{E\,e^{\frac{2\,\pi}{\beta}(t_w\,-\,t^*)\,-\,m\,|x|}}{|x|^{1/2}}
\end{equation}
where $t^*$ is the scrambling time. One can then read off  the Liapunov exponent and the butterfly velocity as $\lambda_L\,=\,\frac{2\,\pi}{\beta}$, and  $v_B^2\,=\,\frac{2\,\pi}{\beta\,m}$, respectively. Now in order to obtain the butterfly velocity, we need to rewrite the function $A(0)$, $B(0)$, $\Phi(0)$ and their derivatives in the original $(t,r,x,y)$ coordinates. Near the horizon,
\begin{eqnarray}
f(r)&\approx & f'(r_{h}) (r-r_{h})+\frac{f''(r_{h})(r-r_{h})^2}{2}+...\\
\nonumber h(r)&\approx & h'(r_{h}) (r-r_{h})+\frac{h''(r_{h})(r-r_{h})^2}{2}...\\
\nonumber A_{t} (r)&\approx &  A'_{t}(r_{h}) (r-r_{h})+\frac{A''_{t}(r_{h})(r-r_{h})^2}{2}...
\end{eqnarray}
we have a relation
\begin{equation}
r^*=\int \frac{1}{\sqrt{h(r)f(r)}} \approx \frac{1}{\sqrt{f'(r_{h})h'(r_{h})}} ln (r-r_{h})
\end{equation}
Thus above relations imply that
\begin{eqnarray}
uv &\approx & -c_{0} (r-rh)\\
A(0) &\approx & -\frac{2}{c_{0}f'(r_{h})}\\
\Phi(0) & \approx & -\frac{A'_{t} (r_{h})}{c_{0} \sqrt{f'(r_{h}) h'(r_{h})}}
\end{eqnarray}
and
\begin{eqnarray}
\nonumber A'(0)&=&\frac{dA(uv)}{d(uv)}|_{u=0}=\frac{dA(uv)}{dr^*} \frac{dr^*}{d(uv)}|_{r_{h}}=\frac{h''(r_{h})}{c_{0}^{2} f'(r_{h}) h'(r_{h})}\\
&\,\,\,\,\,\,\,\ & \Phi'(0)= \frac{A''(r_{h})}{2 c_{0}^{2} \sqrt{f'(r_{h}) h'(r_{h})}} \,\,\,\ ; \,\,\,\ B'(0)=-\frac{2 r_{h}}{c_{0}}
\end{eqnarray}
Considering above equations and using Einstein and Maxwell equations, the effective mass formulas reduces to
\begin{equation}
m^2=r_{h} f'(r_{h})+8 \gamma  r_{h} f'(r_{h})-\frac{8}{3} \gamma  f'(r_{h})^2+\frac{2}{3} \gamma  r_{h}^2 f''(r_{h})^2-4 \gamma  r_{h}^2 f''(r_{h})
\end{equation}
One can now obtain the butterfly velocity as follows:
\begin{equation}
v_B^2\,=\,\frac{\pi\,T\, e^{\frac{\gamma \, q^2}{6 \,r_h^4}}}{ r_h R}
 \label{butterfly velocity}
\end{equation}
where
\begin{equation}
R=1+8 \gamma -\frac{8 \gamma  f'(r)}{3 r}+\frac{2 \gamma  r \left(f''(r)-6\right) f''(r)}{3 f'(r)}.
\end{equation}
for the case $\gamma=0$, it is consistent with the universal formula for butterfly velocities in \cite{1701.05204}.
\section{Holographic stress tensor and the heat current}\label{app4}%
In this section, we are interesting in showing that the radially-conserved bulk current $\mathcal{J}_{Q}$ in (\ref{heat current 1}) matches to the thermal current $Q$, up to a term depending on the time linearly on the $AdS$ boundary at $r \to \infty$. At the boundary, the bulk current $\mathcal{J}_{Q}$ behaves like:
\begin{equation}\label{qw1}
\mathcal{J}_{Q}|_{r\to \infty}=f^{0} \delta g_{tx}'-{f^{0}}' \delta g_{tx}+ \gamma \Big( f_{0} (\mathcal{F}-\zeta^{1})\delta g_{tx}'-{f_{0}}' (\mathcal{F}-\zeta\zeta^{1}) \delta g_{tx}-f^{0} (\mathcal{F}'-2 {\zeta^{1}}') \delta g_{tx}\Big)|_{r\to \infty}-\mu \mathcal{J}
\end{equation}
in above relation, we have used the value of $A_{t}$ at infinity, i.e., the chemical potential, $ A_{t}(\infty)=\mu$. We now focus on the boundary conserved currents through the derivation of the holographic stress tensor. For this purpose, one needs to construct a holographically renormalised on-shell action ($S_{ren}$) \cite{0112119} by adding some terms to the bulk action (\ref{action0}).
\begin{equation}
S_{ren}=S_{bulk}+S_{GH}+S_{ct}
\end{equation}
where the second term is the Gibbons-Hawking term,
\begin{equation}
S_{GH}=2\int d^{3}x \sqrt{-\gamma} K
\end{equation}
which is required for a well defined variational problem with Dirichlet boundary conditions, where $\gamma$ is the determinant of the induced metric $\gamma_{\mu \nu}$ at the boundary and $K$ is the trace of the extrinsic curvature, $K^{\mu \nu}=\nabla^{\mu} n^{\nu}$ where $n_{\mu}$ is the normal vector of the boundary. The last term is counterterms \cite{1409.8346},
\begin{equation}
S_{ct}=\int d^{3}x \sqrt{-\gamma} \Big[-4+\mathcal{R}[\gamma]+\frac{1}{2} \sum_{i=1}^{2} \gamma^{\mu \nu} \nabla_{\mu} \phi_{i} \nabla_{\nu} \phi_{i}\Big]
\end{equation}
which is required to cancel out the UV divergence. Note that in our case the metric on the boundary is flat, so $\mathcal{R}$ vanishes. The renormalized stress tensor $T_{ren}^{\mu \nu}$ and the covariant current $J^{\mu}$ can then be calculated by:
\begin{equation}
T^{\mu \nu}=\frac{2}{\sqrt{\gamma}}\frac{\delta S_{ren}}{\delta \gamma_{\mu \nu}}=-2 \Big[K^{\mu \nu}-K \gamma^{\mu \nu}+2 \gamma^{\mu \nu} +\frac{1}{2} \sum_{i=1}^{2} (\nabla^{\mu} \phi_{i} \nabla^{\nu} \phi_{i}+ \gamma^{\mu \nu} \nabla_{\rho} \phi_{i} \nabla^{\rho} \phi_{i})\Big]
\end{equation}
\begin{equation}
J^{\mu}= \frac{1}{\sqrt{-\gamma}} \frac{\delta S_{ren}}{\delta A_{\mu}}=-n_{\rho}F^{\rho \mu} +4 \gamma C^{\nu \mu \rho \sigma} F_{\rho \sigma} n_{\nu}
\end{equation}
Considering the perturbation (\ref{perturbation of metric and maxwel}) about the black brane, one can obtain elements of the stress tensor as follows:
\begin{equation}
T^{xx}= -\frac{4}{r^2}+\sqrt{\frac{f}{h}} \Big[2+r \frac{h'}{h}\Big]
\end{equation}
\begin{equation}
T^{tx}=\frac{1}{r^2 h}\Big[-4 \delta g_{tx} (t,r)+\sqrt{f} \Big(2 \delta g_{tx} (t,r)+\partial_{r} \delta g_{tx} (t,r)\Big)\Big]
\end{equation}
where a prime denotes the derivative respect to $r$. Now, we can find that
\begin{equation}\label{tt2}
r^2 \sqrt{h} \Big(h T^{tx}-\delta g_{tx}(t,r) T^{xx}\Big)= \sqrt{\frac{f}{h}} \Big(h \partial_{r}\delta g_{tx} (t,r)-h' \delta g_{tx} (t,r)\Big)
\end{equation}
 According to the especial linearised perturbation (\ref{perturbation of metric and maxwel}), we are able to obtain the time-dependent source for $T^{xt}$ as follows:
\begin{equation}\label{TT1}
T^{tx}=\frac{1}{r^2 h}\Big[-4 \delta g_{tx} (r)+\sqrt{f} \Big(2 \delta g_{tx} (r)+\partial_{r} \delta g_{tx} (r)\Big) \Big]-\mathcal{C} t T^{xx}\equiv T^{tx}_{0}-\mathcal{C} t T^{xx}
\end{equation}
where $T^{tx}_{0}$ denotes the time-independent part of stress tensor. Regarding (\ref{TT1}), it is easy to prove that all of the time-dependence terms in (\ref{tt2}) drop out and therefore we finally have:
\begin{equation}
r^2 \sqrt{h} \Big(h T^{tx}_{0}-\delta g_{tx}(r) T^{xx}\Big)= \sqrt{\frac{f}{h}} \Big(h \partial_{r}\delta g_{tx} (r)-h' \delta g_{tx} (r)\Big)
\end{equation}
Evaluating both sides of the above expression on the boundary( $r\to \infty$) up to linear order in $\gamma$, we arrive at:
\begin{equation}\label{trr1}
r^5 T^{tx}=f^{0} \delta g_{tx}'-{f^{0}}' \delta g_{tx}+ \gamma \Big( f_{0} (\mathcal{F}-\zeta^{1})\delta g_{tx}'-{f_{0}}' (\mathcal{F}-\zeta^{1}) \delta g_{tx}-f^{0} (\mathcal{F}'-2 {\zeta^{1}}') \delta g_{tx}\Big)|_{r\to \infty}
\end{equation}
It should be noted that in the large-r limit the metric functions shift to their asymptotic forms $f^0 \sim r^2$ and $\zeta^{1} \sim \mathcal{F} \sim 0$. We have also  $\delta g_{tx} (r) \sim 1/r$, $T^{\mu \nu}\sim 1/r^5$, and $J^{\mu} \sim 1/r^3$ at $r \to \infty$. On the other hand, one can show that $J^{x}$ matches the electric current $\sqrt{-\gamma} \mathcal{J}$ on the boundary exactly.
Therefore, form (\ref{trr1}), up to a term depending on the time linearly, we can conclude that
\begin{equation}
Q=r^5 T^{tx}+\mu r^3 J^{x}
\end{equation}
At linear order in the perturbation, it yields:
\begin{equation}
Q=f^{0} \delta g_{tx}'-{f^{0}}' \delta g_{tx}+ \gamma \Big( f_{0} (\mathcal{F}-\zeta^{1})\delta g_{tx}'-{f_{0}}' (\mathcal{F}-\zeta^{1}) \delta g_{tx}-f^{0} (\mathcal{F}'-2 {\zeta^{1}}') \delta g_{tx}\Big)|_{r\to \infty}-\mu \mathcal{J}
\end{equation}
Therefore, it is obvious that the radially-conserved Noether current (\ref{qw1}) matches precisely to the thermal current on the boundary.
\section{Weyl coupling bound in the presence of two axions}\label{app5}%
The $\gamma$ bound for the schwartzchild black hole with Weyl correction in 4 dimensions is computed in \cite{1010.0443} which is $-1/12<\gamma<1/12$. Two of the authors in \cite{Mansoori:2016zbp} computed the $\gamma$ bounds for Lifshitz black holes with Weyl correction for arbitrary Lifshitz $z$ in any dimension which in the case of $z=1$ and $d=4$, the solution reduces to the results of \cite{1010.0443}. In this paper and in the presence of the two axions we have a hairy black hole with
\begin{equation}\label{Metric M}
f(u)=\frac{r_h^2}{u^2} (1-u^3-\frac{u^2 k^2}{2 r_h^2}(1-u)),
\end{equation}
 where $u=r_h/r$. Note that we ignore the corrections come from Maxwell terms just like the \cite{1010.0443} which easily means taking $q=0$. Now in order to compute the bound we followed the same recipe that have applied in \cite{Mansoori:2016zbp} and \cite{1010.0443}. Finally the potentials  $V_0(u) $ and $ W_0(u)$ which were introduced in \cite{Mansoori:2016zbp} and \cite{1010.0443}, after expanding around $u=0$ and $u=1$, take the following forms:
 \begin{eqnarray}\label{WV}
&& V_0=1+\frac{(4 \gamma -1) k^2 u^2}{2 r_h^2}+ \mathcal{O}(u^3),\nonumber \\
 &&V_0=\frac{ \left(6 r_h^2-k^2\right) \left(-4 \gamma  k^2+12 \gamma  r_h^2+3 r_h^2\right)}{-16 \gamma  k^2 r_h^2+48 \gamma  r_h^4-6 r_h^4}(u-1)\nonumber \\
 &&+\frac {\left (-32\gamma^2 k^6 +
     3\gamma (96\gamma - 17) k^4 {r_h}^2 +
     9\left (-96\gamma^2 + 52\gamma + 1 \right) k^2 {r_h}^4 +
     27 (32 (\gamma - 1)\gamma -
        1) {r_h}^6 \right)} {\left (8\gamma k^2 {r_h} + (3 -
         24\gamma) {r_h}^3 \right)^2}(u - 1)^2\nonumber \\
 &&+ \mathcal{O}((u-1)^3),\nonumber \\
&&W_0=1-\frac{(4 \gamma +1) k^2 u^2}{2 r_h^2}+ \mathcal{O}(u^3),\nonumber \\
 &&W_0=\frac{ \left(6 r_h^2-k^2\right) \left(-8 \gamma  k^2+24 \gamma  r_h^2-3 r_h^2\right)}{-8 \gamma  k^2 r_h^2+24 \gamma  r_h^4+6 r_h^4}(u-1)\nonumber \\
 &&+ \frac {\left (-32\gamma^2 k^6 +
     3\gamma (96\gamma + 25) k^4 {r_h}^2 -
     9\left (96\gamma^2 + 68\gamma - 1 \right) k^2 {r_h}^4 +
     27 (8\gamma (4\gamma + 5) -
        1) {r_h}^6 \right)} {\left (4\gamma k^2 {r_h} -
      3 (4\gamma + 1) {r_h}^3 \right)^2}(u - 1)^2\nonumber \\
      &&+ \mathcal{O}((u-1)^3).
 \end{eqnarray}
For checking the causality in the dual CFT, we have to consider the following constraints on the expansion of the effective potentials at the boundary:
 \begin{eqnarray}\label{Ptent1}
  V_0(u \longrightarrow 0)<1, \,\,\,\,\,   W_0(u \longrightarrow 0)<1
 \end{eqnarray}
In addition, these effective potentials, $V_0(u)$ and $W_0(u)$, show bound states with negative energies, which correspond to unstable quasi-normal modes in the bulk theory. For stability, we request that the energy has to be positive in all directions for a consistent CFT. For this aim, we have to consider the following constraints on the expansion of $V_0(u)$ and $W_0(u)$ near $u = 1$.
\begin{eqnarray}\label{Ptent2}
  V_0(u \longrightarrow 1)>0,\,\,\,\,\   W_0(u \longrightarrow 1)>0
  \end{eqnarray}
As it is clear from equations (\ref{WV}), the bound depends on $k$ and $r_h$. In the incoherent limit $ k\longrightarrow \sqrt{6} r_h$, these equations reduce to the following equations.
 \begin{eqnarray}\label{WV1}
&& V_0=1+3 (4 \gamma -1) u^2+ \mathcal{O}(u^3),\nonumber \\
 &&V_0=-\frac {3 (4\gamma - 1) } {8\gamma + 1} (u - 1)^2+ \mathcal{O}((u-1)^3),\nonumber \\
&&W_0=1-3 (4 \gamma +1) u^2+ \mathcal{O}(u^3),\nonumber \\
 &&W_0=-\frac {3 (8\gamma + 1) } {4\gamma - 1} (u - 1)^2+ \mathcal{O}((u-1)^3).
 \end{eqnarray}
Now according to the relations (\ref{Ptent1}), (\ref{Ptent2}) and equations (\ref{WV1}) we can investigate the $\gamma$ bound as follows:
\begin{equation}\label{gamma bound}
-\frac{1}{8}< \gamma <\frac{1}{4}.
\end{equation}
It is surprising that the above bound on the Weyl coupling can also be obtained by imposing positivity conditions on the butterfly velocity (\ref{vb}) and the conductivity (\ref{sb}). The same behavior for another model has been studied in \cite{Gouteraux:2016wxj}.

\end{document}